\documentclass[%
 reprint,
 amsmath,amssymb,
 aps,
]{revtex4-2}
\usepackage{amsmath}
\usepackage{physics}
\usepackage{amssymb}
\usepackage{graphicx}
\usepackage{float}
\usepackage{tabularx}
\usepackage{xcolor}
\usepackage{graphicx}
\usepackage{dcolumn}
\usepackage{bm}

\usepackage{hyperref}

\usepackage{ulem}

\begin{document}

\preprint{APS/123-QED}

\title{Rate Function Modelling of Quantum Many-Body Adiabaticity}
\author{Vibhu Mishra}
\email{vibhu.mishra@uni-goettingen.de}
\author{Salvatore R. Manmana}%
\email{salvatore.manmana@theorie.physik.uni-goettingen.de}
\author{Stefan Kehrein}
\email{stefan.kehrein@theorie.physik.uni-goettingen.de}
\affiliation{%
 Institute for Theoretical Physics, Georg-August-Universität Göttingen, Germany
}%




\date{\today}

\begin{abstract}

The quantum adiabatic theorem is a fundamental result in quantum mechanics, 
with a multitude of applications, both theoretical and practical.
Here, we investigate the dynamics of adiabatic processes for quantum many-body systems 
by analysing the properties of observable-free intensive quantities.
In particular, we study the adiabatic rate function $f(T, \Delta \lambda)$ in dependence of the ramp time $T$,
which gives us a complete 
characterization of the many-body adiabatic fidelity as a function of $T$ and the strength of the parameter displacement $\Delta \lambda$.
$f(T, \Delta \lambda)$ quantifies the deviation from adiabaticity for a given process and therefore allows us to control and define the notion of adiabaticity in many-body systems. 
First we study $f(T, \Delta \lambda)$ for the 1D Transverse field Ising model and the Luttinger Liquid, both of which are quadratic systems and therefore allow us to look at the thermodynamic limit.
For ramps across gapped phases, we relate $f(T, \Delta \lambda)$ to the transition probability of the system and for ramps across a gapless point, or gapless phase we relate it to the excitation density of the relevant quasiparticles.
Then we investigate the XXZ model which allows us to see the qualitative features that survive when interactions are turned on.
Several key results in the literature regarding the interplay of the thermodynamic and the adiabatic limit are obtained as inferences from the properties of $f(T, \Delta \lambda)$ in the large $T$ limit.
\end{abstract}

\maketitle


\section{\label{sec:level1}Introduction}


The quantum adiabatic theorem, first proposed by Born and Fock \cite{born1928beweis}, is a statement about the time evolution under slow parameter changes of a system initially prepared in one of the eigenstates. 
If we have a Hamiltonian $\mathcal{H}(\lambda(t))$ which is a function of a time dependent parameter $\lambda(t)$, then this defines a family of eigenstates $\ket{m_{\lambda}}$ for each value of  $\lambda$. 
We start with an initial state $\ket{m_{\lambda_i}}$ corresponding to the parameter $\lambda_i$ and change the parameter to $\lambda_f$ over the ramp time $T$ to obtain the time evolved initial state $\ket{\Phi_m(T)} = \hat{\mathcal{U}}(T)\ket{m_{\lambda_i}}$. 
Then the theorem states that the quantum adiabatic fidelity, which is defined as 
\begin{eqnarray}
    \mathcal{F}(T)
    := |\bra{m_{\lambda_f}} \ket{\Phi_m(T)}|^2 \, ,
\end{eqnarray}
for slow enough driving behaves as
\begin{eqnarray}
    \mathcal{F}(T) 
    = 1 - \mathcal{O}(T^{-2})\,.
\end{eqnarray}
This means that for slow driving, the time evolved initial state points towards the corresponding final instantaneous eigenstate of $\mathcal{H}(\lambda(T))$, and that transitions to other energy levels are suppressed. 
The main prerequisite for the application of the theorem is the absence of level crossings. 
The notion of "slow enough" is quantified by the minimum of the energy gap $\Delta E$ along the trajectory, i.e. we require $T \gg \hbar/\text{min}(\Delta E)$. 
In this paper we focus only on ground state adiabaticity and set $\hbar = 1$ throughout. 





It is interesting to note that, while the derivation of the theorem was introduced for single particle systems, 
the applications of the theorem range from relativistic field theories to condensed matter physics.
The adiabatic theorem provides the conceptual foundation for Landau's Fermi Liquid theory \cite{landau1957theory}  and the Gell-Mann Low theorem \cite{PhysRev.84.350}. 
For these cases we assume that such an adiabatic process, in principle, is possible, 
even though the time scales, over which such slow changes take place, would scale with the system size. 
However, this is not a problem 
since here the theorem is used more as a formal device to connect the properties of the eigenstates of the unknown interacting systems with those of the well understood non-interacting ones. 

It is on the practical side, when it comes to adiabatic state preparation (and this can involve a multitude of fields ranging from cold atomic gases \cite{Lewenstein_2007}, adiabatic quantum computation \cite{farhi2000quantum,RolandQuantumAdiabatic}, topological charge pumping \cite{ThoulessTransport} etc.), that it becomes important to have a quantitative handle on the question, as to how slowly a process should be performed to remain within a prescribed bound of adiabaticity. 
In these situations we would like the many-body adiabatic fidelity $\mathcal{F}_N(T)$ for the entire system to deviate as little as possible from unity.


The goal of this paper is to study processes involving time dependent parameter changes in quantum many-body systems by analysing the properties of the adiabatic rate function (which depends on $T$ and the parameter displacement $\Delta \lambda = \lambda_f - \lambda_i$), which we define as
\begin{equation} 
    f(T, \Delta \lambda) := -\lim_{N \rightarrow \infty} \frac{1}{N} \ln (\mathcal{F}_N(T, \Delta \lambda))\, ,
    \label{eq:ratefunction}
\end{equation}
where $N$ is the number of particles.
We evaluate the rate function for the 1D Transverse Field Ising Model (TFIM), the Luttinger Liquid (LL) and the XXZ model.
We will show how the knowledge of the adiabatic rate function can be used to obtain complete information about the many-body fidelity of the corresponding process for sufficiently large $N$ and arbitrary ramp times.
For ramp duration $T=0$, $f(T=0)$  gives us the fidelity following a sudden quantum quench. 
In the large $T$ limit, $f(T, \Delta \lambda)$ gives us the fidelity for an adiabatic process. 
More specifically, $f(T, \Delta \lambda)$ will give us the corresponding value of $T$ for which a prescribed value of the many-body fidelity $\mathcal{F}_N(T, \Delta \lambda)$ is achieved. 
We will also see how the qualitative properties of $f(T, \Delta \lambda)$ change depending on the phases and the path through the phase diagram, which we choose for the ramp, and the implications it has for the approach to adiabaticity in terms of the dependence of adiabatic time scales with system size.
 
The remainder of the paper is structured as follows:
In Sec.~\ref{sec:3RF_Definition} we motivate the definition of the adiabatic rate function.
In Sec.~\ref{sec:Prior results} we go through several well known key results on the problem of adiabatic quantum many-body dynamics and alternative approaches to adiabaticity in many-body systems. 
In Sec.~\ref{sec:RF_Non_Interacting} we evaluate the rate functions for TFIM and LL and try to understand the results we obtain.

In Sec.~\ref{sec:XXZ_RF} we will study the rate functions for the XXZ spin chain and see at least in the gapped phase, how the qualitative behavior does not change for an interacting system.
We conclude with a summary and outlook in Sec.~\ref{sec:summary}.
In the appendices we go through the technical calculations.

\section{Adiabatic Rate Function}
\label{sec:3RF_Definition}
\subsection{Definition}
\label{subsec:3a}
We start with a many-body Hamiltonian with a time dependent parameter $\mathcal{H}(\lambda(t))$. Each value of $\lambda$ defines a set of instantaneous eigenstates $\ket{n_\lambda}$. We choose the initial and final parameters $\lambda_i$ and $\lambda_f$ respectively and the ramp time $T$ which decides the duration of the process.
We choose a simple linear ramp,
\begin{equation}
\label{eq:parameter}
\lambda(t) = \lambda_i + \frac{t}{T}(\lambda_f - \lambda_i),
\end{equation}
where time $t \in [0,T]$.

At $t=0$, we are in the ground state of  $\lambda_i$, which is $\ket{0_{\lambda_i}}$. 
Then we perform the unitary time evolution $\hat{\mathcal{U}}(T)$ generated by $\mathcal{H}(\lambda(t))$, and ask how well does the time evolved state vector point towards the ground state corresponding to the final parameter value $\lambda_f$.

The adiabatic many-body fidelity for a system with $N$ particles is given by 
\begin{equation}
    \mathcal{F}_N(T) = |\bra{0_{\lambda_f}}\hat{\mathcal{U}}(T)\ket{0_{\lambda_i}}|^2.
\end{equation}
We posit the following ansatz for this fidelity
\begin{equation}
\label{eq:ansatz}
\mathcal{F}_N(T)  \equiv \exp(-N \cdot   f_N(T, \lambda_i, \lambda_f)) \, ,
\end{equation}
from which we obtain the adiabatic rate function ($\lambda_i, \lambda_f$ removed for clarity of notation henceforth)
\begin{eqnarray}
\label{eq:f_N(T)}
    f_N(T) = - \frac{1}{N} \ln \left(\mathcal{F}_N(T) \right) \, .
\end{eqnarray}
$f_N(T)$ is the quantity that can be obtained via numerics. 
The definition becomes useful if $f_N(T)$ has a well defined thermodynamic limit for the global parameter quenches that we discuss in this paper. 
We define this limit as the rate function 
\begin{eqnarray}
    f(T) \equiv \lim_{N \rightarrow \infty} f_N(T) \, .
\end{eqnarray}
 If $f_N(T)$ converges sufficiently quickly then for large $N$ we can evaluate $\mathcal{F}_N(T)$ using 
 \begin{equation}
\label{eq:ansatz_f(T)}
\mathcal{F}_N(T)  = \exp(-N \cdot   f(T)) \, .
\end{equation}
In practice, this means that the rate function can be used to obtain all the information about the many-body overlaps for large $N$ and arbitrary ramp times $T$.

We follow the discussion in Ref.\cite{bachmann2017adiabatic} and motivate the ansatz in Eq.~\eqref{eq:ansatz} heuristically by looking at many-body product states: 
for example, for two quantum spin-1/2 particles, whose directions are off by a small angle $\theta$, their direct overlap would be
\begin{equation}
\label{eq:Direct_Overlap}
|\bra{\psi + \theta}\ket{\psi}|^2 = 1 - \mathcal{O}(\theta) \approx \exp(-\mathcal{O}(\theta)) \, .
\end{equation}

If we have $N$ such pairs of decoupled spins, due to the tensor product structure, the overlap of the entire quantum state is just the product of individual overlaps:
\begin{equation}
\label{eq:N_Body_Overlap}
|\bra{\psi + \theta}\ket{\psi}_N|^2 = \prod_i \exp(-\mathcal{O}(\theta)) = \exp(-N\mathcal{O}(\theta)) \, .
\end{equation}
We point out that the equation above works only for global parameter changes (i.e., all the angles $\theta$ in the entire many-body system are changed).
For local perturbations, the situation can be very different, and the $N$ dependence of the direct overlap in Eq.~\eqref{eq:N_Body_Overlap} changes its form. 
For example, in Anderson's original work on the orthogonality catastrophe \cite{anderson1967infrared}, he studied the effect of impurities in metallic systems and obtained overlaps, which decay algebraically with system size as $N^{-\epsilon}$, where $\epsilon$ quantifies the strength of the impurity.

Another heuristic way of understanding this ansatz is to note that a the Hilbert space  for $N$ spin-$1/2$ particles in a spin chain has size $2^N$. 
If we decompose any randomly chosen state $\ket{\phi}$ in terms of its basis states $\ket{i}$
then generically we would expect $|\bra{i} \ket{\phi}|^2 \sim 2^{-N}$ to be valid for the whole trace to add up to $1$.
Hence for large $N$, in terms of the rate function, we can write $|\bra{i} \ket{\phi}|^2 \sim e^{-Nf(i,\phi)}$.


Next, we want to include the explicit time evolution in the expressions. 
In the aforementioned case this corresponds to some physical process, which aligns the spins of state $\ket{\psi}$ with those of $\ket{\psi + \theta}$, where it takes a time $T$ to change the angle by $\theta$.
Here, we simply absorb the $T$-dependence in the exponential, 
\begin{equation}
|\bra{\psi + \theta} \mathcal{\hat{U}}(T) \ket{\psi}_N|^2 =  \exp(-N\mathcal{O}(\theta, T)) \, .
\end{equation}
For this process, the rate function will be
\begin{eqnarray}
    f(T) = \mathcal{O}(\theta, T) \, .
\end{eqnarray}


Working in the quantum many-body regime presents us with an interesting complication, which is the presence of a multitude of quantum phases and critical points in the parameter space.
Critical points or gapless phases imply that the energy gaps vanish in the thermodynamic limit.
The adiabatic dynamics sensitively depends on the behavior of energy gaps and consequently on whether the initial and final parameters lie entirely within the same phase or have a critical point in between them.  
As we will see, the specifics of the process in the parameter space will lead to qualitative changes in the behavior of the properties of $f(T)$.

$f(T)$ has a nice physical interpretation as it is a non-negative, intensive quantity. 
Therefore, it is well defined in the thermodynamic limit and provides a quantitative measure of deviation from adiabaticity for the entire range of ramping times, for small $T$ (fast ramps) to large $T$ (adiabatic limit), independent of the system size. 
Knowledge of $f(T)$ can be used to quantitatively evaluate the fidelity for any system size $N$, of the corresponding physical process (from $\lambda_i$ to $\lambda_f$). 
On the theoretical side, analyzing the properties of $f(T)$ in the large $T$ limit,  along with the ansatz of our many-body overlap in Eq.~\eqref{eq:ansatz}, will allow us to obtain the results that we will discuss in Sec.~\ref{sec:Prior results}.


\subsection{Distinction from Loschmidt Echo}

The definition of the adiabatic rate function is very similar to that of the Loschmidt Echo so here we explicitly make a distinction between the two.

If we start with some initial state $\ket{\psi_i}$ (usually the ground state of some Hamiltonian $\ket{0_i}$) then the return amplitude or the Loschmidt amplitude for this state under some generic time evolution is simply the overlap between the time evolved state with the initial state 
\begin{equation}
    \bra{0_i}\hat{U}(t)\ket{0_i} = e^{-Nf(t)},
\end{equation}
from which we obtain $f(t)$ as the Loschmidt rate function \cite{heyl2019dynamical}.
Due to its formal similarity with the boundary partition function for imaginary values of $t$, the Loschmidt rate function was shown to possess non-analytic behavior (in real $t$) for quenches across different phases for both integrable and non-integrable models \cite{Heyl_Polkovnikov_Kehrein_DQPT, DQPT_Non_Integrable}.

The Loschmidt rate function is also investigated in the context of Quantum Speed Limits (QSL) where the problem of interest is to investigate the minimum time it takes for a time evolved system to become orthogonal to its original initial state \cite{QSL}.

The clear difference between the Loschmidt rate function and the adiabatic rate function is that we are interested in the fidelity between the time evolved initial state with respect to the final state and not the initial state and are interested in the behavior of this fidelity with respect to the \textit{ramp time} (or the inverse speed) $T$.

\section{Properties of many-body adiabatic processes}
\label{sec:Prior results}
Adiabatic processes in many-body settings have been studied in great detail, see, e.g., Refs.\cite{de2010adiabatic, dziarmaga2005dynamics, lychkovskiy2017time, polkovnikov2008breakdown, polkovnikov2005universal, dziarmaga2010dynamics}. 
In this section, we highlight the most relevant results (we use in the following the notation applied in this paper). 

Lychkovskiy \textit{et al.} \cite{lychkovskiy2017time} present a lower bound on the ramp time $T$ (inverse of the driving speed $\Gamma_N$ in their notation) for an adiabatic process with very modest inputs in terms of the scaling properties of the the many-body ground states. 
These scaling properties are characterized by parameters, which are denoted as $C_N$ and $\delta V_N$, and  are introduced as follows: 

$C_N$ quantifies the strength of the orthogonality catastrophe. 
It is obtained by exploring the dependence of many-body overlaps for different values of some system parameter $\lambda$ in the large $N$ limit,
\begin{eqnarray}
    \mathcal{C}(\lambda) 
    \equiv |\bra{\Phi_\lambda} \ket{\Phi_0}|^2 
    = \exp(-C_N \lambda^2).
\end{eqnarray}
For example, for impurity models $C_N \sim \ln(N)$, while for global parameter changes (i.e the situations considered in this paper) one obtains $C_N \sim N$.

$\delta V_N \equiv \sqrt{\langle\hat{V}^2\rangle_0-\langle\hat{V}\rangle_0^2}$ is the uncertainty in the driving potential with respect to the ground state where $\left.\hat{V} \equiv \frac{\partial \hat{H}}{\partial \lambda}\right|_{\lambda=0}$.
From these one obtains bounds for the ramp time needed to observe adiabatic behavior for the $N$ body state, namely
\begin{equation}
T \geq \frac{2 C_N}{\delta V_N}\,.
\end{equation}
For example, for a gapped system where we perform a global parameter change adiabatically, we have $\delta V_N \sim \sqrt{N}$ and $C_N \sim N$, and therefore
\begin{equation}
\label{eq:Oleksander_result}
T \geq 2\sqrt{N} \, .
\end{equation} 

In \cite{QSL_RF}, Suzuki and Takahashi investigate QSL bounds under counter-diabatic (CD) driving, for both the Loschmidt rate function and the adiabatic rate function which they define as 
\begin{equation}
    |\bra{\psi_{\text{ad}}(t)}\ket{\psi(t)}| \sim e^{-Ng_{\text{ad}}(t)},
\end{equation}
where $\ket{\psi(t)}$ is the time evolved initial state and $\ket{\psi_{\text{ad}}(t)}$ is the appropriate instantaneous eigenstate at the particular time.
This rate function measures the deviation from adiabaticity during the specific process and agrees with the definition given in Eq.~\eqref{eq:ansatz} specifically for end point of the process $t=T$.
They plot $g_{\text{ad}}(T)$ vs $T$ for the Transverse Field Ising Model and find a monotonically decaying behavior which is what we would expect as smaller  $g_{\text{ad}}(T)$ implies being closer to adiabaticity.
It is this specific problem that we focus on in our present work.

In \cite{polkovnikov2005universal}, Polkovnikov focuses on adiabatic processes in the vicinity of quantum critical points and relates the density of excitations $n_{\rm ex}$ (number of excited states upon the total number of states) to the critical exponents $\nu, z$, the ramp time $T$  (again the inverse of the driving speed $\delta$ in their notation), and the dimensionality $d$ as
\begin{equation}
n_{\text{ex}} \propto T^{-d \nu/(1+z\nu)}\,. 
\end{equation} 
For the case of the one dimensional transverse field Ising model, for which we have $\nu = d = z = 1$, we get
\begin{eqnarray}
\label{eq:TFIM_Excitation_Density}
    n_{\text{ex}} \propto  \frac{1}{\sqrt{T}}  \, ,
\end{eqnarray}
which matches the result of Ref.\cite{dziarmaga2005dynamics} derived using the Landau-Zener formula.

In another work \cite{de2010adiabatic}, De Grandi and Polkovnikov  find that the density of excitations of noninteracting quasiparticles within a gapless phase having dispersion relation $\epsilon_k = c(\lambda)k^z$ turns out to be
\begin{equation}
n_{\text{ex}} \propto T^{-d/z} \, .
\end{equation} 
For a Tomonaga-Luttinger liquid, $d=z=1$ and we find $n_{\text{ex}} \propto 1/T$.

Here we would like to point out that for the case of global parameter changes, (which means $C_N \sim N$) all of these results in different cases emerge naturally as properties of the rate function $f(T)$ discussed in this paper, in the large $T$ limit. 
For non-interacting gapped systems we will show $f(T)$ to be directly related to the transition probabilities of individual subsystems whereas for gapless systems or critical points, $f(T)$ measures the density of quasiparticle excitations for any given ramp time $T$.

A completely separate direction is taken by Bachmann \textit{et al.} \cite{bachmann2017adiabatic} who base their analysis on observable dependent quantities. They work with a different definition of adiabaticity, which they define as the convergence of the expectation values of some local operator $\hat{O}$, with respect to the time evolved initial state $| \psi(s)\rangle$, and the instantaneous eigenstate at the final parameter value $| \Omega_s\rangle$ ($C$ is some non-universal constant independent of the system size and $\varepsilon$ is some small parameter)
\begin{equation}
\left|\langle\psi(s) | \hat{O} | \psi(s)\rangle-\langle\Omega_s | \hat{O} | \Omega_s\rangle\right| \leq C \varepsilon \,.
\end{equation}
They show that the adiabatic condition so defined  becomes independent of the system size. 
In our work, in contrast, we stick to analysing the behavior of the wave-function overlaps by studying the rate function.



\section{$f(T)$ for non-interacting systems}

\label{sec:RF_Non_Interacting}
\subsection{Gapped systems}
\label{subsec:RF_Gapped}

We make use of the Born-Fock result, to find $f(T)$ for gapped non-interacting system.
For an individual gapped subsystem, the Born-Fock result shows that for $T$ much larger than the minimum of the inverse gap size, the following relation holds
\begin{equation}
\label{eq:1_Body_Ansatz}
 \mathcal{F}(T)= |\bra{0_{\lambda_f}}\hat{\mathcal{U}}(T)\ket{0_{\lambda_i}}|^2 = 1 - \mathcal{O}(T^{-2}).
\end{equation}
If we have a tensor product of such systems (the energy gaps need not be the same) the many-body fidelity is simply the product of individual fidelities
\begin{equation} 
\begin{split}
\mathcal{F}_N(T) =& \prod_{k=0}^N |\bra{0^k_{\lambda_f}}\hat{\mathcal{U}}_k(T)\ket{0^k_{\lambda_i}}|^2 \\ =& \prod_{k=0}^N(1 - \mathcal{O}_k(T^{-2})) \approx \prod_{k=0}^N \exp(-\mathcal{O}_k(T^{-2})),
\end{split}
\end{equation}
where $k$ is just an index for individual subsystem (but will usually correspond to the quasimomentum).

The product of exponentials becomes a sum of the exponents and gives us
\begin{equation}
\mathcal{F}_N(T) = \exp(-N \cdot \frac{1}{N}\sum_k \mathcal{O}_k(T^{-2})),
\end{equation}
from which we can immediately write down $f_N(T)$ as
\begin{equation}
\label{eq:BF_RF}
f_N(T) =  \frac{1}{N}\sum_k \mathcal{O}_k(T^{-2}) \propto \frac{1}{T^2}.
\end{equation}
Since each individual term in the summation is simply a transition probability which lies between $0$ and $1$, therefore the summation is well defined in the thermodynamic limit.
We see that for this simple problem, the adiabatic rate function is just the average of the transition probabilities of the individual sub-systems.



\subsection{Transverse Field Ising Model}

\label{subsec:TFIM}
The TFIM is one of the most important models in condensed matter physics because it is the simplest microscopic model that exhibits second order quantum phase transition.
In 1D, a sequence of Jordan-Wigner and Fourier transformations \cite{TFIM} maps the model to a collection of non-interacting two state systems which can then be diagonalized via Bogoliubov transformation.
This allows us to test our ideas for large system sizes and also makes our discussion in Sec.~\ref{subsec:RF_Gapped} directly relevant to interpret the results that we find.

The specific Hamiltonian that we start with is
\begin{equation}
    H(t) = -J\sum_{j=1}^N \sigma_z^j\sigma_z^{j+1} - h(t) \sum_{j=1}^N  \sigma_x^j.
\end{equation}
It furnishes us with two gapped phases, the paramagnetic or disordered phase for $|h/J| > 1$ and the symmetry broken ferromagnetic phase for $|h/J| < 1$ and positive $J$ (anti-ferromagnetic for negative $J$).
For $|h/J| = 1$ we have a gapless critical point.
We work with $J = 1$ and positive $h$ throughout which means ferromagnetic long range order for $h < 1$.

After Jordan-Wigner and Fourier transformation we have
\begin{equation}
    H(t) = \sum_k H_{k}(t),
\end{equation}
where (setting $J = 1$)
\begin{equation}
\begin{split}
    H_{k}(t) = 2(h(t) &-  \cos k)\left(c_{k}^{\dagger} c_{k} - c_{-k} c_{-k}^{\dagger}\right) \\ &- 2  \sin k\left(i  c_{k}^{\dagger} c_{-k}^{\dagger}-i  c_{-k} c_{k}\right).
\end{split}
\end{equation}
In the fermionic basis we work with anti-periodic boundary conditions for the real space fermionic operators, which means that $c_1 = -c_{L+1}$. 
This sets the allowed $k$ values as $k= \pm \frac{(2 n-1) \pi}{L}, \text { with } n=1, \cdots, \frac{L}{2}$ \cite{TFIM}.

Following the Bogoliubov transform we obtain
\begin{equation}
    H_k(t) = \epsilon_k(t) \gamma_k^\dagger(t) \gamma_k(t),
\end{equation}
where the Bogoliubov operators $\gamma_k(t)$ (which are explicitly time dependent in this case by definition) annihilate the ground state at the instant $t$ and provide us with the instantaneous eigenbasis for the entire process for a given $h(t)$.
The eigenenergies of the individual $k$ modes are can be calculated to be
$\epsilon_{k}(t) = 2 \sqrt{(\cos k-h(t))^{2} + \sin ^{2} k}$.

For $h > 1$, the Bogoliubov Hamiltonian gives the unique paramagnetic ground state but for $h < 1$ we obtain the positive parity superposition of the symmetry broken states as our true ground state.

For the process of interest, the external magnetic field $h(t)$ is changed linearly from $h_i$ to $h_f$ over a ramp time $T$.
The ground state is a product state over the quasi-momentum modes of the Bogoliubov vacuua $\ket{0_{h_i}^k}$
\begin{equation}
    \ket{0_{h_i}} = \prod_k \ket{0_{h_i}^k}
\end{equation}
The adiabatic fidelity is given by ($\mathcal{\hat{U}}_k(t)$ is the unitary time evolution generated by $H_k(t)$)
\begin{equation}
   \begin{split}
       |\bra{0_{h_f}} \mathcal{\hat{U}}(T) \ket{0_{h_i}}|^2 &= \prod_k |\bra{0_{h_f}^k} \mathcal{\hat{U}}_k(T) \ket{0_{h_i}^k}|^2 \\ &= \exp(-Nf_N(T)),
   \end{split}
\end{equation}
where we have used Eqs. \eqref{eq:Direct_Overlap} and \eqref{eq:N_Body_Overlap}.
This immediately gives us  
\begin{equation}
\label{eq:TFIM_RF_Discreet}
    f_N(T) = -\frac{1}{N} \sum_k \ln(|\bra{0_{h_f}^k} \mathcal{\hat{U}}_k(T) \ket{0_{h_i}^k}|^2).
\end{equation}

It is this quantity that we evaluate by numerically solving the time dependent Schrödinger equation
using the method of instantaneous eigenstates \cite{griffiths2018introduction, de2010adiabatic} (the relevant differential equation being Eq.~\eqref{eq:DiffEQSimpler}), which gives \textit{exact} results for quadratic systems, and is discussed in  \ref{subsec:Method_of_Instantaneous_Eigenstates}.
This is because for each instant of time $t$, we can diagonalize the Hamiltonian, write down the eigenstates and the energy spectrum and define an instantaneous eigenbasis which is non-degenerate.

More specifically, for a particular $k$ mode, we evaluate the adiabatic fidelity by solving the differential equation in Eq.~\eqref{eq:DiffEQSimpler} for $m=0$ which corresponds to ground state fidelity and plug it in Eq.~\eqref{eq:TFIM_RF_Discreet}.

For large $N$, Eq. \eqref{eq:TFIM_RF_Discreet} tends to a well defined limit (we justify this in Sec.\ref{subsec:RF_Gapless})
\begin{equation}
\label{eq:Exact_Ising}
\begin{split}
    f(T) &= \lim_{N \rightarrow \infty}f_N(T)
    \\
    & = - \int_{-\pi}^{\pi} dk \ln(|\bra{0_{h_f}^k} \mathcal{\hat{U}}_k(T) \ket{0_{h_i}^k}|^2).
\end{split}  
\end{equation}
This is because for ramps across gapped phases, each term in the sum over $k$ is a small deviation from unity and the discussion in Sec.\ref{subsec:RF_Gapped}, in particular Eq.~\eqref{eq:BF_RF} applies.
For ramps across the critical point, the assumption of small deviation from unity breaks down for certain $k$ modes and we need to be more careful in our analysis which is carried out in Sec. \ref{subsec:RF_Gapless}. 
It is due to the breakdown of this assumption that we get qualitatively different behavior between $f(T)$ for gapped vs. gapless cases.

We also need to address the question whether Eq.~\eqref{eq:Exact_Ising} is well defined for small $k$ when the energy gap closes (but when there is still no level crossing). 
In such a case, for a given $T$, for small $k$, we are effectively performing a quantum quench protocol.
For such $k$ modes we have to look at the adiabatic fidelity in the quenched limit i.e $\lim T \rightarrow 0$ which gives (for small $\Delta h = h_f - h_i$)
\begin{align}
     |\bra{0_{h_f}^k} \mathcal{\hat{U}}_k(T) \ket{0_{h_i}^k}|^2 \nonumber 
    & \geq |\bra{0_{h_f}^k}  \ket{0_{h_i}^k}|^2
    \\
    & \approx 1 - (\Delta h)^2 \,.
\end{align}
Taking the logarithm of the above expression is unproblematic which means that taking the continuum limit is justified and therefore $f(T)$ is well defined for both gapped and gapless systems that we are studying.

\subsubsection*{Ramps within the gapped phase}

\begin{figure}
    \centering
    \includegraphics[scale=0.19]{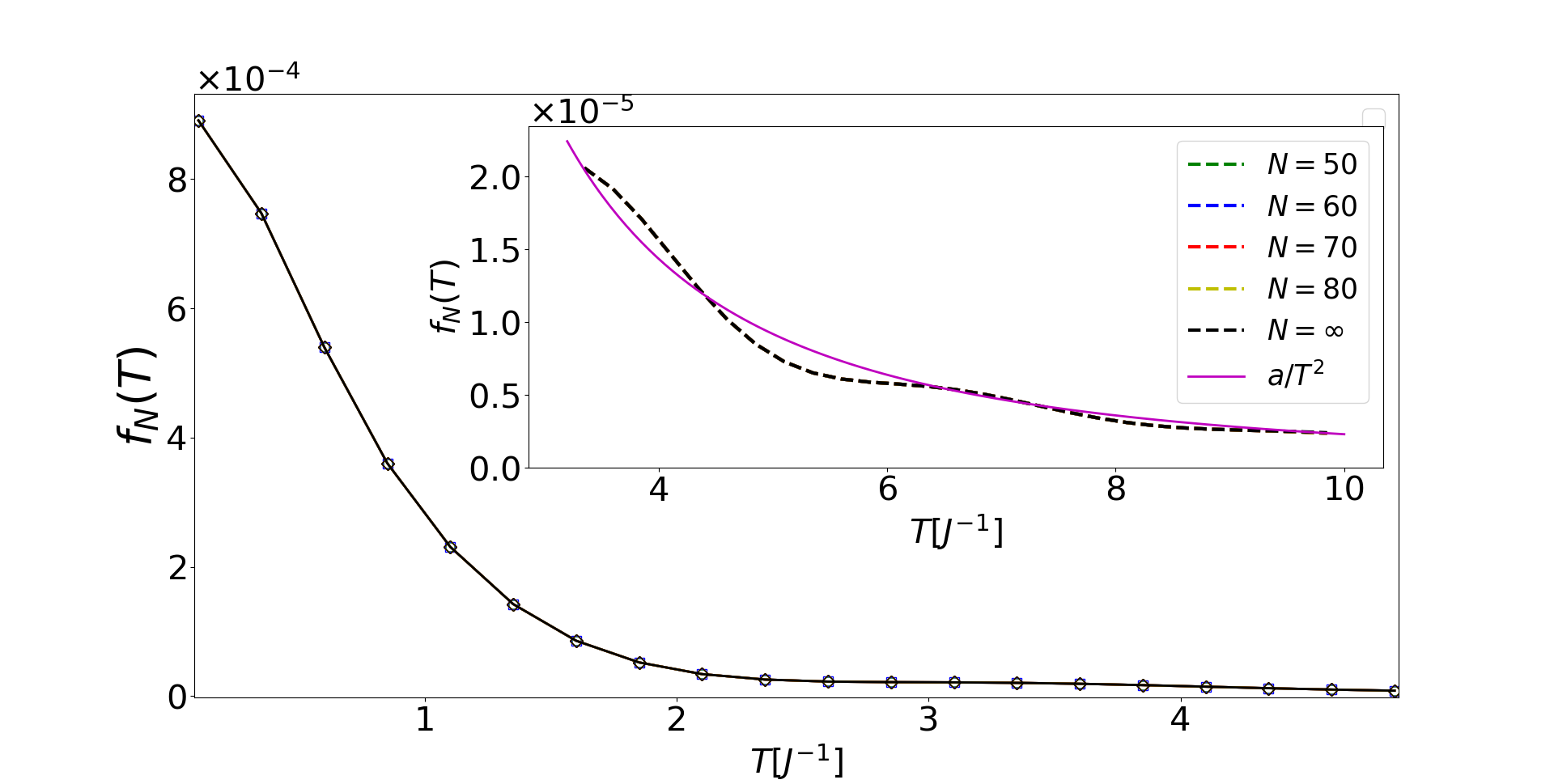}
    \vfill
    \caption{$f_N(T)$ for ramps within the paramagnetic phase of the TFIM. $h(t)$ ramps from $1.4$ to $1.6$.  
    The inset is a zoom into the large $T$ region and shows the finite-size data in comparison to a curve fit of the form $a/T^2$. 
    As can be seen, even for relatively modest system sizes, $f_N(T)$ looks independent of $N$.
    Finite size extrapolation ($N = \infty$) is carried out using a quadratic fit function $f_N(T) = g + h/N + i/N^2$ where  $g$ provides the value of $f(T)$. }
    \label{fig:TFIM_RF_Gapped}
\end{figure}

\begin{figure}
	\centering
\includegraphics[scale=0.19]{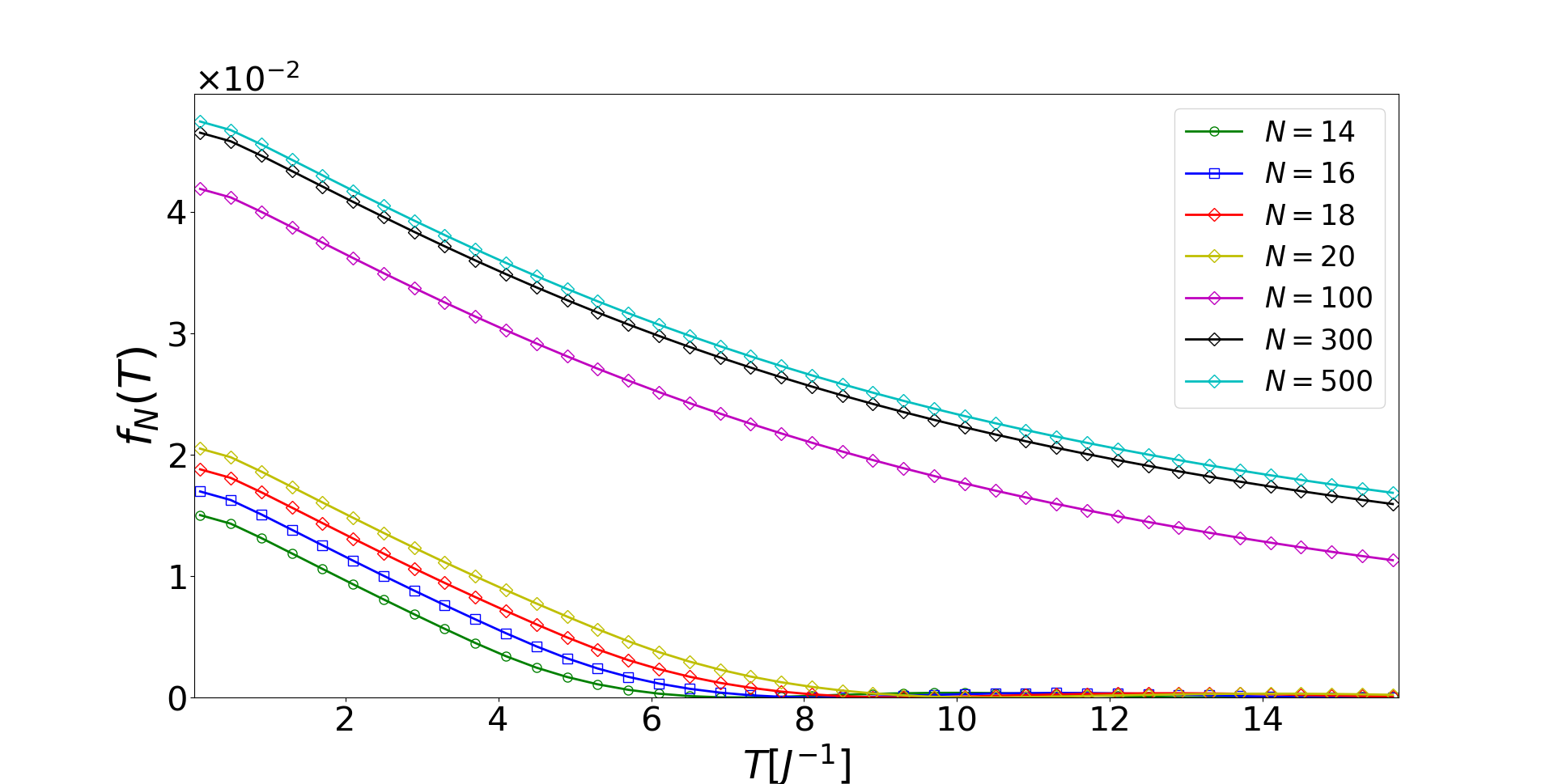}
\vfill

	\caption{Ramps across the Ising critical point for small $N$. $h(t)$ ramps from $1.1$ to $0.9$. 
 We need to go to extremely large $N$ to curb the finite size effects.
 }
	\label{fig:TFIM_Critical_Small_N}
\end{figure}

In Fig.~\ref{fig:TFIM_RF_Gapped} we see that $f(T)$ shows monotonic decay for small $T$ followed by an oscillatory decay beyond a certain $T$.
The oscillations themselves are highly non-universal (and can be seen in cases as simple as spin precession, see App.~\ref{App:B}) but, inspired by the discussion in Sec.~\ref{subsec:RF_Gapped}, we can check for the $1/T^2$ algebraic decay and find a nice agreement.
We also see that $f_N(T)$ quickly reaches the thermodynamic limit.
For ramps across the critical point, we see in Fig.~\ref{fig:TFIM_Critical_Small_N} that this is simply not the case and there are extremely strong finite-size effects.


\begin{figure}
    \centering
    \includegraphics[scale=0.19]{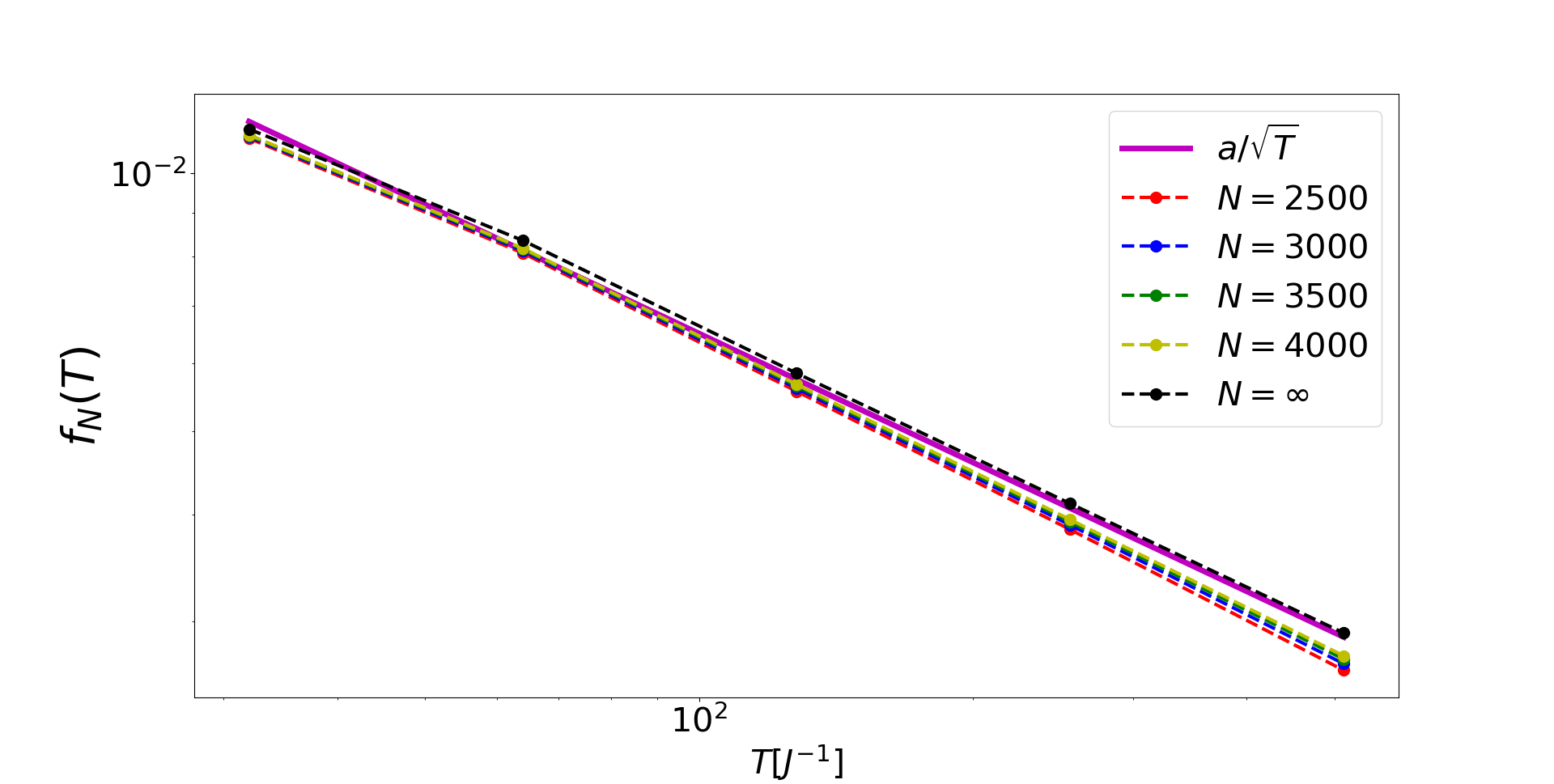}
    
    \vfill

    \caption{A log-log plot for $f_N(T)$ for ramps across the Ising critical point. 
    $h(t)$ ramps from $1.1$ to $0.9$. 
    The curve fit is carried out by $a/\sqrt{T}$ where $a$ is the curve fitting parameter. 
 Finite size extrapolation ($N = \infty$) is carried out using a quadratic fit function $f_N(T) = g + h/N + i/N^2$ where  $g$ provides the value of the rate function at $N = \infty$.}
    \label{fig:TFIM_Critical_f(T)}
\end{figure}

\subsection{Gapless systems}
\label{subsec:RF_Gapless}

For gapless systems, since the inverse energy gaps are proportional to the system size, for $T \ll N$ the ansatz in the right side of Eq.~\eqref{eq:1_Body_Ansatz} breaks down as the energy gap requirement of the adiabatic condition is not satisfied.
In Eq.~\eqref{eq:BF_RF} the transition probability $\mathcal{O}_k(T^{-2})$ is replaced by $\mathcal{G}_k(T)$ which is also decaying in $T$ but whose form we do not know.

The key point to consider is that for any given $T$, there is a collection of $k$ modes which are excited more strongly than others.
For the case of TFIM, ramps across the critical point map to a collection of Landau-Zener systems.
Since the minimum energy gap for any $k$ mode is proportional to $k$ itself, we can use the Landau-Zener formula to obtain the transition probability as ($\alpha$ contains all the other numerical prefactors of $\pi, 2, \hbar$ etc.)
\begin{equation}
\label{eq:Landau_Zener}
    \mathcal{G}_k(T) \sim e^{-\alpha k^2 T}.
\end{equation}
For a given $T$, this gives us the notion of a critical $k$ mode which we call $k_C(T) \propto 1/\sqrt{T}$ such that modes lying between $k = -k_C(T)$ and $k = k_C(T)$ are strongly excited and therefore have a large $\mathcal{G}_k(T)$. 
The adiabatic rate function (using Eq.~\eqref{eq:Exact_Ising}) for such a situation can then be approximated as 
\begin{equation}
\begin{aligned}
    f(T) &= - \int_{-\pi}^{\pi} dk \ln(1-\mathcal{G}_k(T)) \\
    &\approx - \int_{-k_c(T)}^{k_c(T)} dk \ln(1-\mathcal{G}_k(T)).
\end{aligned} 
\end{equation}

A simple rewriting in terms of $\Tilde{k} = k\sqrt{T}$ gives us (the integrand becomes independent of $T$ except for the prefactors)
\begin{equation}
    f(T) \propto \frac{1}{\sqrt{T}},
\end{equation}
which connects us back to Eq.~\eqref{eq:TFIM_Excitation_Density} and can be seen in Fig.\ref{fig:TFIM_Critical_f(T)}.
It suggests that for ramps across Ising critical point, $f(T)$ is just a measure of the density of excitations for any given $T$ or more generally
\begin{equation}
\label{eq:RF_Critical_Point}
    f(T) \propto n_{\text{ex}}(T).
\end{equation}

In the next section we test Eq.~\eqref{eq:RF_Critical_Point} for the Luttinger liquids.

\subsection{Luttinger Liquid}

\label{subsec:TLL}

LL theory is a universal low energy model that substitutes Landau's Fermi Liquid theory for 1D metallic wires \cite{giamarchi2003quantum, 1DFL, Refermionization}.
The original fermionic Hamiltonian is re-expressed in terms of operators that create or destroy particle-hole excitations which are bosonic in nature.
The model predicts several features distinct from regular Fermi Liquids.
The surprising feature of LL is that even interacting fermionic models get mapped to non-interacting bosonic Hamiltonians with the Luttinger parameters being controlled by the parameters of the original fermionic model.

The specific bosonized Hamiltonian we work with is \cite{iucci2009quantum} (the set of allowed $q$ values are $q= \pm \frac{(2 n-1) \pi}{L}, \text { with } n=1, \cdots, \frac{L}{2}$)
\begin{equation} 
\label{eq:H_TLL}
\begin{split}
H_{\text{LL}} &= \sum_q H_q(t) \\  &= \sum_q |q| \left( (\omega + \Delta(t) ) b^{\dagger} _q b _q + \frac{\Delta(t)}{2}  \left[b _q b_{-q}+b^{\dagger}_q b^{\dagger}_{-q}\right] \right)\,.
\end{split}
\end{equation}


We introduce $\alpha_q = \frac{1}{\sqrt{2}} (b_q + b_{-q})$ and $\beta_q = \frac{i}{\sqrt{2}} (b_q - b_{-q})$ and re-express the previous Hamiltonian in terms of these new operators to obtain (note that $q$ only takes positive values from here on)
\begin{equation}
H_{\text{LL}}  = \sum_{q>0} H_q(t) + H_{-q}(t) = \sum_{q>0} H_q^{\alpha}(t) + H_q^{\beta}(t)\,,
\end{equation}
where ($a = \alpha, \beta$ labels the particle)
\begin{equation}
\label{eq:TLL_H_Used}
H^a_q(t) = q \left((\omega + \Delta(t) ) a^{\dagger}_q a_q + \frac{\Delta(t)}{2}\left[(a_q)^2 + (a_q^{\dagger})^2 \right]   \right)\,.
\end{equation}
Since $[\alpha_q, \beta_q^\dagger] = 0$ for all $q$, the Hilbert space parts into two sectors (the individual $q$ modes of $\alpha$ and $\beta$ particles).

We change the value of $\Delta(t)$ from $\Delta_i$ to $\Delta_f$ linearly over a ramp time $T$.
The initial ground state is just a product state of the ground states of Eq.~\eqref{eq:TLL_H_Used}
\begin{equation}
    \ket{0_{\Delta_i}} = \prod_{a = \alpha, \beta} \prod_q \ket{0_{\Delta_i}^q}_a
\end{equation}

The adiabatic fidelity is given by ($\mathcal{\hat{U}}_q^a(t)$ is the unitary time evolution generated by $H^a_q(t)$)
\begin{equation}
   \begin{split}
       |\bra{0_{\Delta_f}} \mathcal{\hat{U}}(T) \ket{0_{\Delta_i}}|^2 &= \prod_{a = \alpha, \beta} \prod_q |\bra{0_{\Delta_f}^q} \mathcal{\hat{U}}_q^a(T) \ket{0_{\Delta_i}^q}_a|^2 \\ &= \exp(-Nf_N(T))
   \end{split}
\end{equation}
This again gives us  
\begin{equation}
\label{eq:LL_RF_Discreet}
    f_N(T) = -\frac{1}{N} \sum_{a = \alpha, \beta}\sum_q \ln(|\bra{0_{\Delta_f}^q} \mathcal{\hat{U}}_q^a(T) \ket{0_{\Delta_i}^q}_a|^2),
\end{equation}
which in the continuum limit takes the form (the integration is over half of the Brillouin zone now)
\begin{equation}
\label{eq:Exact_LL}
\begin{split}
    f(T) &= \lim_{N \rightarrow \infty}f_N(T)
    \\
    & = - \sum_{a = \alpha, \beta}\int_0^{\pi} dq \, \ln(|\bra{0_{\Delta_f}^q} \mathcal{\hat{U}}_q^a(T) \ket{0_{\Delta_i}^q}_a|^2).
\end{split}  
\end{equation}

Again we can make use of the method of instantaneous eigenstates (outlined in the next subsection) to evaluate the expression in the above equation.
The eigenstates and the spectra have been evaluated in Appendix[\ref{App:C}].

While the above expression is exact, easy to evaluate and plot numerically, it is too schematic to give any insights into the obtained results.
Since the system is gapless, $T$ never fulfills the adiabatic condition therefore the discussion in Sec.\ref{subsec:RF_Gapped} does not apply.

To gain qualitative insights we make a series of approximations and then compare them with the exact result in Eq.~\eqref{eq:Exact_LL}.
We start with re-expressing Eq.~\eqref{eq:Exact_LL} in terms of the transition probability $\mathcal{G}_q(T)$ for a particular $q$ mode and anticipating the result to be similar to the case of ramp across the Ising critical point, we make a further truncation of the integral (sum over particle flavor $a$ is suppressed for clarity)
\begin{equation}
\label{eq:f(T)_trunctation}
\begin{aligned}
    f(T) &= - \int_0^{\pi} dk \ln(1-\mathcal{G}_q(T)) \\
    &\approx - \int_0^{q_c(T)} dq \ln(1-\mathcal{G}_q(T)).
\end{aligned} 
\end{equation}

Following the discussion from Sec.~\ref{subsec:RF_Gapless}, for a particular value of $T$, there is a collection of modes ranging from $0$ to $q_c(T)$, where $q_c(T)$ is the upper bound on the number of modes which are strongly excited during the process. 
Strong excitations correspond to large transition probabilities ($\mathcal{G}_q(T)$ is large) which means that the behavior of $f(T)$ is dominated by these strongly excited modes.

The density of excitations for LL for given ramp time $T$ is \cite{de2010adiabatic} 
\begin{eqnarray}
    q_c(T) \propto 1/T\, .
\end{eqnarray}

\begin{figure}
    \centering
    \includegraphics[scale=0.195]{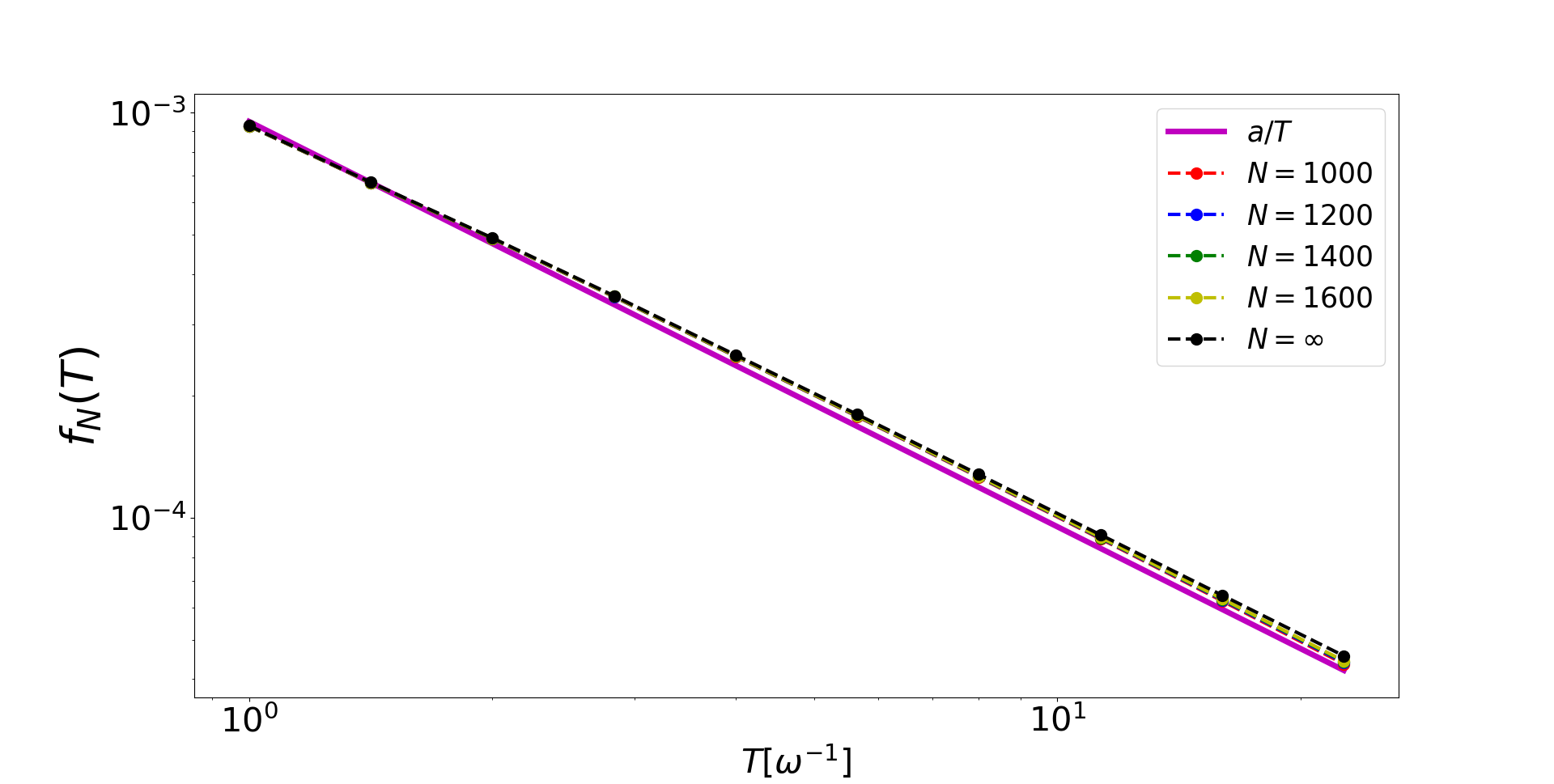}

    \vfill

    \caption{$f_N(T)$ for the Luttinger Liquid. $J_z(t)$ ramps from $-0.1$ to $0.1$. 
    The curve fit is carried out by $a/T$ where $a$ is the curve fitting parameter. 
    This validates the result obtained in Eq.~\eqref{eq:LL_1/T}.
    Finite size extrapolation ($N = \infty$) is carried out using a quadratic fit function $f_N(T) = g + h/N + i/N^2$ where  $g$ (the intersection of the curve with the y-axis) provides the value of $f(T)$.}
    \label{fig:LL_f(T)}
\end{figure}

Re-expressing the integral in terms of a rescaled variable $Tq = q'$, gives us ($\mathcal{G}_q(T)$ becomes independent of $T$ when expressed in terms of $q'$, see Eq.\eqref{eq:adia_condition_rescaled})  
\begin{eqnarray}
    \label{eq:LL_1/T}
    f(T) \propto 1/T \, .
\end{eqnarray}
The result is confirmed in Fig.\ref{fig:LL_f(T)}, which is evaluated exactly \textit{without} making the approximation of Eq.~\eqref{eq:f(T)_trunctation}.
This implies that $f(T)$ has the same dependence on $T$ as the density of excitations $n_{\text{ex}}(T)$. 

Eq.~\eqref{eq:LL_1/T} is a direct consequence of the approximation made in Eq.~\eqref{eq:f(T)_trunctation}, which has a very general character and can be applied whenever we have the knowledge of $n_{\text{ex}}(T)$ vs $T$.
The values of $\alpha$ for $n_{\text{ex}}(T) \propto 1/T^\alpha$ have been evaluated for various critical points (for the Ising critical point $\alpha = 1/2$) and can be found in Ref.\cite{dziarmaga2010dynamics}. 
For all these cases, we expect to get $f(T)\propto 1/T^\alpha$.

\subsection{Method of Instantaneous Eigenstates}

\label{subsec:Method_of_Instantaneous_Eigenstates}

Here we provide more details for the evaluation of Eqs. \eqref{eq:Exact_Ising} and \eqref{eq:Exact_LL} by going through the relevant  steps ~\cite{de2010adiabatic, griffiths2018introduction}.

We consider the time-dependent Hamiltonian and for each instant of time $t$, it defines an orthonormal basis set ($n=0$ is the instantaneous ground state), 
\begin{equation}
\hat{H}(t) \ket{n(t)} = E_n(t) \ket{n(t)},
\end{equation}
where for all $t$ we have
\begin{equation}
\bra{m(t)}\ket{n(t)} = \delta_{m,n} \, .
\end{equation}
We can expand any generic quantum state at time $t$ in terms of these instantaneous eigenstates $\ket{n(t)}$, and energies $E_n(t)$  
\begin{equation}
\ket{\Psi(t)} = \sum_n c_n(t) \ket{n(t)} \exp(-i \int_0^t  dt' E_n(t')) \, .
\end{equation} 
If the initial state is some eigenstate of $\hat{H}(t=0)$ labeled by some index $m$, which gives us the initial conditions as
\begin{equation}
\ket{\Psi(t = 0)} = \ket{m(t=0)} \implies c_n(t=0)  = \delta_{m,n} \, .
\end{equation}
The advantage of this framework is that we only need to worry about the behavior of $c_n(t)$, provided all other quantities ($\ket{\psi_n(t)}, E_n(t)$) in principle are known. 
This is the case for all Bogoliubov type models, be it fermionic (Ising model, Kitaev honeycomb model, XY model etc) or bosonic (Luttinger liquids).

The differential equations for the coefficients $c_l(t)$ are 
\begin{equation} 
\label{eq:DiffEq}
\begin{split}
\dot{c}_{l}(t)=&-c_{l}(t)\langle l(t) | \dot{l}(t)\rangle \\ &-\sum_{n \neq l} c_{n}(t) \frac{\langle l(t)|\dot{H}(t)| n(t)\rangle}{E_{nl}(t)} e^{-i  \int_{0}^{t} E_{nl}\left(t^{\prime}\right)   dt^{\prime}},
\end{split}
\end{equation}
where by definition $\ket{\dot{\psi}(t)} \equiv \frac{\partial}{\partial t} \ket{\psi (t)}$ and $E_{nl}\left(t\right) \equiv E_{n}\left(t\right)-E_{l}\left(t\right)$.

The first term of Eq.~\eqref{eq:DiffEq} can safely be ignored because we can always choose a gauge for any generic state $\ket{\psi}$
\begin{equation}
\ket{\psi(t)} = e^{i \theta(t)}\ket{\phi(t)} 
\end{equation}
by choosing $\theta(t)$ such that
\begin{equation}
\bra{\psi(t)} \ket{\dot{\psi}(t)} = i \dot{\theta}(t) + \bra{\phi(t)} \ket{\dot{\phi}(t)} = 0\,.
\end{equation}
The set of coupled differential equations we need to solve are
\begin{equation}
\label{eq:DiffEQSimpler}
\dot{c}_{l}(t)=-\sum_{n \neq l} c_{n}(t) \frac{\langle l(t)|\dot{H}(t)| n(t)\rangle}{E_{nl}(t)} e^{-i \int_{0}^{t}E_{nl}\left(t^{\prime}\right)    dt^{\prime}}.
\end{equation}
A clearer way to express the differential equations is to write them in terms of rescaled time $s = t/T$
\begin{equation} \label{eq:adia_condition_rescaled}
\begin{split}
\frac{d c_l(s)}{ds} = -\sum_{n \neq l} c_{n}(s) \frac{\left\langle l(s)|\frac{d H}{ds}| n(s)\right\rangle}{E_{nl}(s)} e^{-i \int_{0}^{s}E_{nl}(s^{\prime})    T ds^{\prime}}
\end{split}
\end{equation}
and we see how the total ramp time $T$ controls the behaviour of the fidelities via the complex exponential. 

For quadratic systems like TFIM or LL, the above expression can be evaluated numerically by solving the differential equations without any approximations.

For concreteness we look at a particular $k$ mode of the TFIM.
The relevant overlap of interest is $\bra{0_{h_f}^k} \mathcal{\hat{U}}_k(T) \ket{0_{h_i}^k}$.

We expand the time evolved initial state as
\begin{equation}
\label{eq:InstantaneousEigenstateExpansion_TFIM}
    \begin{split}        \mathcal{\hat{U}}_k(t) \ket{0_{h_i}^k} = &c_0^k(t) \ket{0_{h(t)}^k} \exp(-i \int_0^t  dt' E_0^k(t')) \\ 
    &+ c_1^k(t) \ket{1_{h(t)}^k} \exp(-i \int_0^t  dt' E_1^k(t'))
    \end{split}
\end{equation}
$\ket{0_{h(t)}^k}, \ket{1_{h(t)}^k}$ are the instantaneous Bogoliubov ground state and excited state respectively.

At $t=0,$ $h_i = h(0)$ which gives us the initial conditions as $c_0^k(t=0) = 1 \text{ and } c_1^k(t=0) = 0$

At $t=T,$ $h_f = h(T)$ which gives us the required overlap as
\begin{eqnarray}
    \bra{0_{h_f}^k} \mathcal{\hat{U}}_k(T) \ket{0_{h_i}^k} = c_0^k(T) \exp(-i \int_0^T  dt' E_0^k(t')),
\end{eqnarray}
and the rate function therefore is simply
\begin{equation}
f(T) = -\frac{1}{N}\sum_k \ln(|c_0^k(T)|^2). 
\end{equation}

To solve Eq.~\eqref{eq:DiffEQSimpler} we need to evaluate the matrix overlap which for TFIM turns out to be
\begin{equation}
    \frac{\langle 1^k_{h(t)}|\dot{H}_k(t)| 0_{h(t)}^k\rangle}{E_{10}^k(t)} =  \frac{-2i \sin(k)(h_f-h_i)}{4T(\sin^2(k) + (h(t) - \cos^2(k)))} .
\end{equation}

The steps to be followed for LL are almost exactly the same as that of TFIM.
The major difference is that Eq.~\eqref{eq:InstantaneousEigenstateExpansion_TFIM} becomes an infinite sum due to the bosonic nature of the system which has to be truncated for us to be able to perform numerics.

The transition matrix elements for LL (which interestingly are independent of $q$) can be shown to be
\begin{equation}
    \frac{\langle (n+2)^q_{\Delta(t)}|\dot{H}_q(t)| n_{\Delta(t)}^q\rangle}{E_{n+2, n}^q(t)} = \frac{(\Delta_f - \Delta_i)\sqrt{(n+1)(n+2)}}{2 T (\omega(\omega + 2\Delta(t)))} .
\end{equation}



\section{$f(T)$ for Interacting Systems}

\label{sec:XXZ_RF}

Now we focus on the 1D XXZ model. 
It has a rich phase diagram which allows us to look at various regimes and allows us to see which features of $f(T)$ are retained when we are looking at genuinely interacting systems.

The particular Hamiltonian we work with is
\begin{equation}
H(t) = -J_{xy}\sum_{i = 1}^N \left( S^x_i S^x_{i+1} + S^y_i S^y_{i+1} \right) - J_z(t) \sum_{i = 1}^N S^z_i S^z_{i+1} .
\end{equation}
We set $J_{xy} = 1$ and ramp the time dependent parameter $J_z(t)$.
Within the gapped anti-ferromagnetic ($J_z(t) < -1$) and ferromagnetic ($J_z(t) > 1$) phases, we make use of exact diagonalization and use the python package QuSpin \cite{weinberg2017quspin,weinberg2019quspin} which allows us to treat system sizes $N \sim 20$. 

\subsection{Within Gapped Phases}

\subsubsection{Within FM Phase ($J_z(t)>1.0$)}
The ferromagnetic ground state simply consists in all spins pointing in the same direction with the spin-z magnetization $\langle \sigma_z \rangle = \pm1$. 
There is only one single state within the chosen magnetization sector, and since the $H_{\text{XXZ}}$ conserves magnetization symmetry, there is no possibility for the system to make a transition to any other state no matter how fast or slow the driving for $J_z(t)$ is. Since $f(T)$ measures transition probabilities, this means that, for driving within the FM phase, we obtain $f(T) = 0$.

\subsubsection{Deep within AFM Phase ($J_z(t) \ll -1.0$)}
\label{sec:AFM_RF}
The AFM phase presents us with two degenerate ground-states (the two Néel states). 
We restrict ourselves to the positive parity subsector by choosing the symmetric combination of the two degenerate Néel states as our initial ground state. 

\begin{figure}
    \centering
    \includegraphics[scale=0.25]{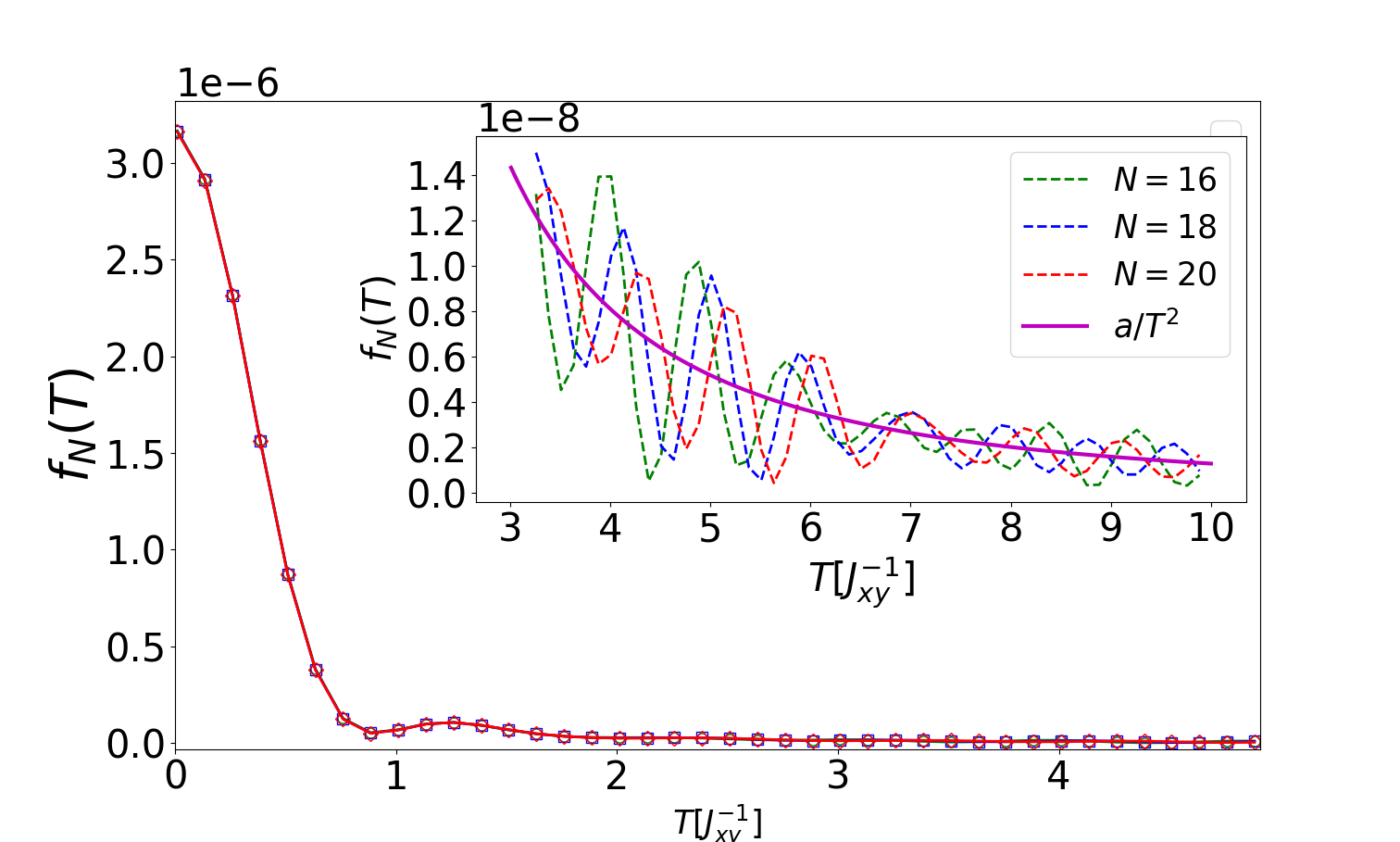}
    \caption{$f_N(T)$ for ramps deep within the AFM phase of the XXZ spin chain. $J_z(t)$ ramps from $-7.4$ to $-7.6$. 
    The inset is a zoom into the large-$T$ region and shows the finite-size data in comparison to a curve fit of the form $a/T^2$. 
    As we can see, the difference between the results for different $N$ is very small, so that they cannot be distinguished in the main plot.}
    \label{fig:AFM_RF}
\end{figure}

Eq.~\eqref{eq:BF_RF} implies that even for this interacting many-body gapped phase the behavior should follow $f(T) \sim 1/T^2$.
In the inset of Fig.~\ref{fig:AFM_RF} we verify this, as shown by the solid line, which is intended as guide to the eye.
The data oscillates around this line, but otherwise it follows this behavior well.


\subsection{Ramps within the Gapless XY Phase ($|J_z(t)| < 1.0$)}
\label{sec:LL_f(T)}
\begin{figure}
	\centering
		\includegraphics[scale=0.25]{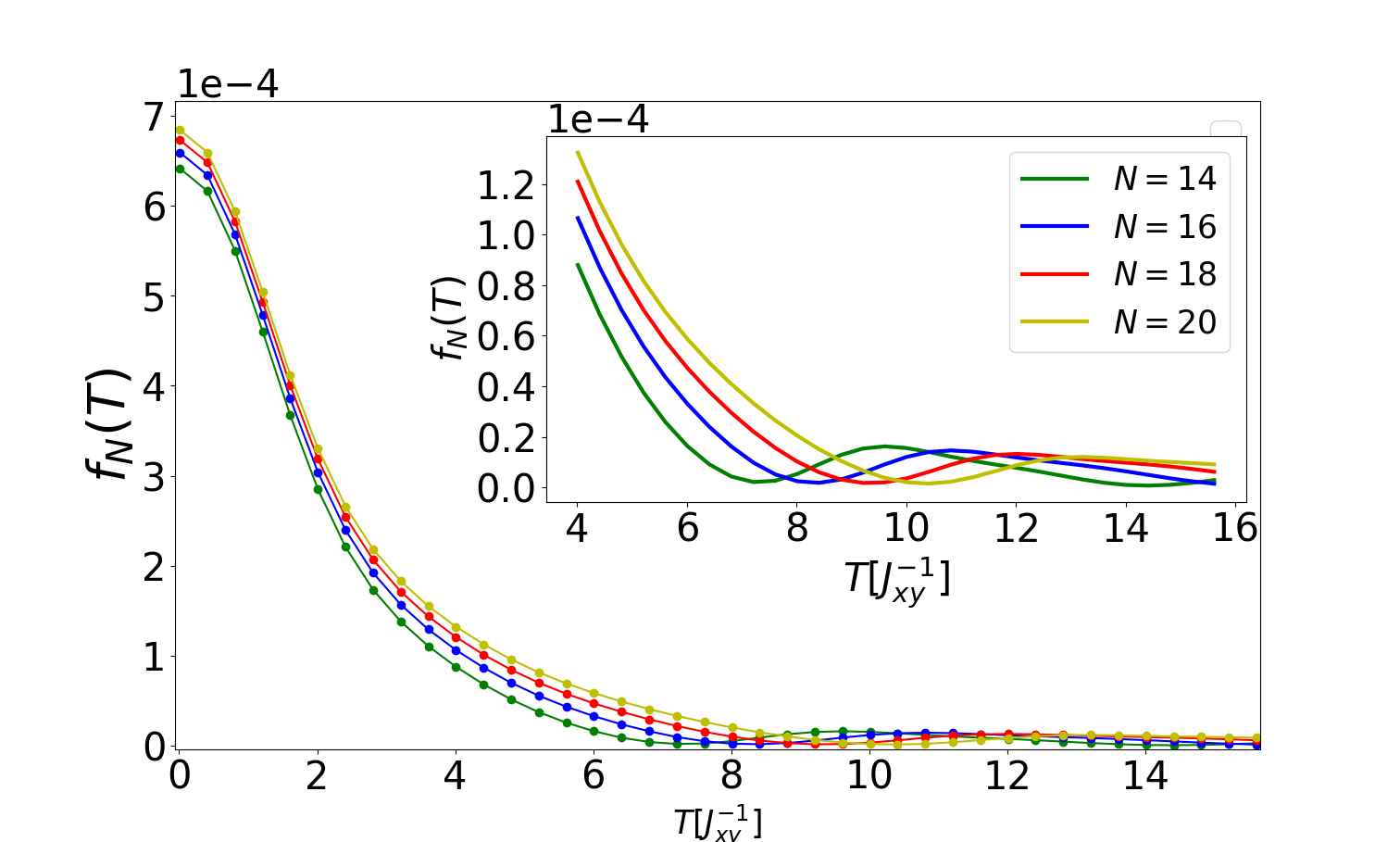}

	\caption{Exact diagonalization results for ramps in the XXZ chain when staying within the XY phase for small $N$. 
    $J_z(t)$ ramps from $-0.1$ to $0.1$. 
    In the inset plot we see that the position of the first minimum moves towards larger values of $T$ when increasing $N$.}
	\label{fig:XXZ_LL}
\end{figure}

In Fig.~\ref{fig:XXZ_LL} we present our results for ramps staying inside the gapless XY phase, i.e. $|J_z(t)|<1 \, \forall t$. 
While it is possible to investigate $f_N(T)$ for the finite systems shown in the Fig. (up to $N=20$ sites), it is now much more difficult to perform an extrapolation to the thermodynamic limit in order to obtain the rat function $f(T)$.
This is further illustrated in the inset of Fig.~\ref{fig:XXZ_LL}, where we see that the position of the first minimum moves to larger $T$ when increasing $N$.
For the TFIM, we observe in Fig.~\ref{fig:TFIM_Critical_Small_N} similar behavior for $N \lesssim 20$.
There, only when treating $N \gtrsim 2000$ sites we are able to obtain $f(T)$. 
This is a system size, which is beyond the scope of the methods used here.
In future work, this can be further analyzed using methods which directly treat infinite systems, like infinite size matrix product states (iMPS)~\cite{SCHOLLWOCK201196,PAECKEL2019167998,Karrasch2013}.
However, we expect the picture obtained from our analysis based on LL in Sec.~\ref{sec:LL_f(T)} to hold also for the XXZ model for ramps staying inside this phase.


\section{Summary and Outlook}
\label{sec:summary}

In this article we investigated the physics of time dependent parameter changes in quantum many-body systems by analyzing the properties of adiabatic rate functions $f(T)$.
We expect our findings to be relevant for various realizations, like cold atomic gases \cite{Lewenstein_2007}, adiabatic quantum
computation \cite{farhi2000quantum}, many-body topological charge pumping \cite{Bertok_FHM}, where it is possible to write down simple model Hamiltonians which describe the system at hand, and allow for the evaluation of $f(T)$.

In Sec.~\ref{subsec:3a}, we started our analysis by noting that for global parameter changes, the many-body fidelity between any two quantum states decays exponentially with system size $N$. 
This implies that increasing the ramp time $T$ for any time-dependent parameter would bring the relevant many-body fidelity closer and closer to unity (which is the adiabatic limit). 
In contrast, for any given $T$, increasing $N$ will diminish these overlaps to zero. 

To study this competition between the thermodynamic and the adiabatic limits systematically, we first motivated an ansatz for the fidelity of a many-body system under a time dependent parameter change and separated the $T$ and the $N$ dependent parts.
This has  allowed us to define the adiabatic rate function $f(T)$ for the process which we have shown to be an intensive quantity that does not depend on $N$.
We have shown how the knowledge of $f(T)$ provides us with the many-body fidelity for any $N$ for the corresponding process both in the small $T$ (quench) and the large $T$ (adiabatic) limit. 
We have shown how the knowledge of $f(T)$ along with our ansatz of many-body fidelity provided us with the notion of an adiabatic time scale $T_N$ which depends on $N$ and insures that on these time scales the entire many-body fidelity evolves close to unity.

In Sec.~\ref{subsec:RF_Gapped}, we evaluated $f(T)$ for the simple case of a collection of gapped non-interacting systems. 
We have found that for $T$ larger than the inverse energy gaps, $f(T)$ is just the average of the transition probabilities of the individual sub-systems and therefore $f(T) \sim 1/T^2$.

In Sec.~\ref{sec:RF_Non_Interacting} we tested our ideas by evaluating $f(T)$ for the TFIM and LL. 
For processes within the gapped phases, we have confirmed that $f(T) \sim 1/T^2$, and that $f(T)$ quickly reaches a well defined thermodynamic limit even for  relatively small $N$. 
For processes in the LL phase and for ramps across the critical point of the TFIM we have found that $f(T) \sim 1/T$ and $f(T) \sim 1/\sqrt{T}$ respectively. 

We have derived these decay exponents by first pointing out the direct relation between $f(T)$, and the density of excitations for these processes, and then relating the excitation density to the ramp time $T$. 
The idea of relating $f(T)$ to the density of quasi-particle excitations is extremely general and is valid for other models which describe the physics in the vicinity of quantum critical points as well.
The key results of this paper are the approximations in Eq.~\eqref{eq:BF_RF} and Eq.~\eqref{eq:RF_Critical_Point}, which are compared with the exact results provided by Eq.~\eqref{eq:Exact_Ising} and Eq.~\eqref{eq:Exact_LL}.

In Sec \ref{sec:XXZ_RF} we have studied the XXZ model using exact diagonalization and for ramps across gapped phase found the expected $f(T) \sim 1/T^2$ and discussed the extremely strong finite size effects in the gapless XY phase which renders a systematic analysis unfeasible within the scope of this article.

Along such lines it would be interesting to investigate $f(T)$ for genuinely interacting systems hopefully using a combination of analytical and numerical methods.
It will also be interesting to search for new qualitative features by studying the behavior of $f(T)$ for systems that do not allow for a quasi-particle description,  e.g., the SYK model~\cite{Rosenhaus_2019}.



\section{Acknowledgments}
\label{sec:acknowledgements}
We thank Marin Bukov for insightful discussions and Rishabh Jha for helpful comments on an earlier version of the manuscript.
We acknowledge financial support by the Deutsche Forschungsgemeinschaft (DFG, German Research Foundation) - 217133147/SFB 1073, projects B03 and B07.

\section{DATA AVAILABILITY}
\label{sec:Data}
All data and simulation codes used to generate the figures and the associated quantities presented in this study is
available at Zenodo upon reasonable request \cite{Data_Set}.

\renewcommand*{\bibfont}{\footnotesize}

\appendix

\section{Oscillatory Behavior of $f(T)$}
\label{App:B}

We start with the spin precession problem from chapter 11 of Ref.\cite{griffiths2018introduction}.
A spin-$1/2$ particle with an energy gap $\Delta$ between the two states is initially pointing along the direction of an external magnetic field. 
The field slowly changes direction by some angle in a long time $T$. 
For the process to be adiabatic, the spin should point along the final direction of the magnetic field and the probability for a spin flip during this process should go to zero for increasing values of $T$.

The transition probability for this process is evaluated to be \cite{griffiths2018introduction}
\begin{eqnarray}
    \chi_{\Delta} \propto \frac{1}{(T \Delta )^2} \sin^2\left( \frac{T \Delta }{2}\right).
\end{eqnarray}
This transition probability corresponds to $\mathcal{O}(T^{-2})$ in Eq.~\eqref{eq:1_Body_Ansatz}.
Now if we have a collection of $N$ such slowly precessing spin-$1/2$ particles with energy gaps $\Delta_i$ (which need not be equal), then we can write down $f(T)$ for this system by making use of Eq.~\eqref{eq:BF_RF}, 
\begin{equation}
    f_N(T) =  \frac{1}{N}\sum_{i=1}^N \chi_{\Delta_i} \propto \frac{1}{T^2} \cdot \frac{1}{N}\sum_{i=1}^{N} \frac{1}{( \Delta_i )^2} \sin^2\left( \frac{T \Delta_i }{2}\right).
\end{equation}

This is just the average of the transition probabilities and we find for this simple problem that $f(T)$ has the form $1/T^2$ times some oscillating part.
The behavior of oscillations depends on the details of $\Delta_i$ but the $1/T^2$ decay is universal.

\section{Instantaneous eigenstates of Luttinger Liquids}
\label{App:C}

We start by completely solving the following toy problem \cite{coleman2015introduction} (i.e., to compute the spectra and the eigenstates) as its results will be used directly for the LL rate functions  in section(\ref{subsec:TLL}):
\begin{equation}
\label{eqC.1}
H=(\omega + \Delta)\left(a^{\dagger} a+\frac{1}{2}\right)+\frac{1}{2} \Delta\left(a^{\dagger} a^{\dagger}+a a\right) \,.
\end{equation}
We define new bosonic operators, with the goal of diagonalizing the above Hamiltonian, as
\begin{equation}
\begin{aligned}
b &=u a+v a^{\dagger} \\
b^{\dagger} &=u a^{\dagger}+v a \, ,
\end{aligned}
\end{equation}
with $u^2-v^2 = 1$. 
This immediately gives us
\begin{equation}
\begin{aligned}
a &=u b-v b^{\dagger} \\
a^{\dagger} &=u b^{\dagger}-v b \, .
\end{aligned}
\end{equation}
We introduce the parameter $k$ defined as
\begin{equation}
\label{eq:Def_k}
k \equiv \frac{\omega }{\Delta}\,.
\end{equation}
Rewriting the Hamiltonian in terms of the new operators we get
\begin{align}
H &= (\omega + \Delta)\left((u b^{\dagger}-v b)(u b-v b^{\dagger}) +  \frac{1}{2}\right) \\& + \frac{1}{2}\Delta \left((u b-v b^{\dagger})(u b-v b^{\dagger})+(u b^{\dagger}-v b)(u b^{\dagger}-v b)  \right) \, .
\end{align}
Imposing the condition that 
\begin{equation}
H=\tilde{\omega}\left(b^{\dagger} b+\frac{1}{2}\right)
\end{equation}
immediately yields the following two conditions:
\begin{equation} 
(\omega + \Delta)(u^2 + v^2) - 2\Delta uv =  \tilde{\omega}  
\end{equation}
and
\begin{equation}
 \frac{\Delta}{2}(u^2+v^2)  -uv(\omega + \Delta) =  0 \, .
\end{equation}
Using hyperbolic functions to parametrize $u$ and $v$ 
\begin{equation}
u = \cosh \theta, \: v = \sinh \theta
\end{equation}
gives us equation
\begin{equation}
\tanh \theta + \coth \theta = 2(k+1)\, ,
\end{equation}
where $k$ is taken from Eq.~\eqref{eq:Def_k}.

Imposing the requirement that $|v/u| = |\tanh \theta| < 1$ gives us the following solutions to this quadratic equation: 
\begin{equation}
    \begin{aligned}
        \tanh \theta &= k+1 - \sqrt{k(k+2)}, \: k>0
        \\
        \tanh \theta &= k+1 + \sqrt{k(k+2)}, \: k< -2\,.
    \end{aligned}
\end{equation}
The value of $k$ cannot lie between $(-2,0)$ because then the quantity inside the square root would be negative.

Using hyperbolic function identities we can now get the values of $u,v$ in terms of $k$.
For $k>0$
\begin{equation}
u = \sqrt{\frac{k+1 + \sqrt{k(k+2)}}{2\sqrt{k(k+2)}}}, v= \sqrt{\frac{k+1-\sqrt{k(k+2)}}{2\sqrt{k(k+2)}}}
\end{equation}
and for $k<-2$
\begin{equation}
u = \sqrt{\frac{k+1 - \sqrt{k(k+2)}}{2\sqrt{k(k+2)}}}, \, v= \sqrt{\frac{k+1+\sqrt{k(k+2)}}{2\sqrt{k(k+2)}}} \, .
\end{equation}
The eigenenergies are
\begin{equation}
\label{eq:LL_Energy}
E_n = (n+1/2)\tilde{\omega} = (n+1/2)\sqrt{\omega(\omega + 2\Delta)} \, .
\end{equation}

Now we have to demonstrate that the boson paired coherent state defined as
\begin{equation}
\label{eq:LL_GS}
|\tilde{0}\rangle = \mathcal{N} \exp(-\alpha\left(a^{\dagger} a^{\dagger}\right))|0\rangle\,,
\end{equation}
is the ground state of the total interacting Hamiltonian ($|0\rangle$ is the ground state for the non-interacting part).
$\mathcal{N}$ is the normalization which we evaluate in App.~\ref{App:D}.

The value of $\alpha$ is obtained by imposing the condition that the new ground state is annihilated by the operator $b$, 
\begin{equation}
\label{eq:LL_GS_Constraint}
    \begin{aligned}
        b|\tilde{0}\rangle&=\left(u a+v a^{\dagger}\right)|\tilde{0}\rangle
        \\
        &=\left(u a+v a^{\dagger}\right)\exp(-\alpha\left(a^{\dagger} a^{\dagger}\right))|0\rangle= 0 \, .
    \end{aligned}
\end{equation}
To evaluate this we need the commutator $\left[a, e^{-\alpha a^{\dagger} a^{\dagger}}\right]$. 
After expanding the exponential we get
\begin{equation}
\left[a, e^{-\alpha a^{\dagger} a^{\dagger}}\right] = \sum_{n=1}^{\infty} \frac{(-\alpha)^n}{n!}\left[a, (a^{\dagger} a^{\dagger})^n\right] \, .
\end{equation}
To get the commutator on the R.H.S we start with
\begin{equation}
\left[a, a^{\dagger} \right] = 1,
\end{equation}
to obtain
\begin{equation}
\left[a, a^{\dagger}a^{\dagger} \right] = 2a^{\dagger}\,.
\end{equation}
We now use this to obtain
\begin{equation}
\left[a, (a^{\dagger}a^{\dagger})^n \right] = n(a^{\dagger}a^{\dagger})^{n-1}\left[a, a^{\dagger}a^{\dagger} \right] = 2n(a^{\dagger})^{2n-1}\,.
\end{equation}
The step above is correct only because the commutator $\left[a, a^{\dagger}a^{\dagger} \right] = 2a^{\dagger}$ trivially commutes with the operator $a^{\dagger}a^{\dagger}$.

Plugging back in Eq.~\eqref{eq:LL_GS} we get
\begin{equation}
    \begin{aligned}
        &\sum_{n=1}^{\infty} \frac{(-\alpha)^n}{n!}\left[a, (a^{\dagger} a^{\dagger})^n\right]
        \\
        &= \sum_{n=1}^{\infty} \frac{(-\alpha)^n}{n!}2n(a^{\dagger})^{2n-1} 
        \\
        &=-2\alpha a^\dagger \sum_{n=1}^{\infty} \frac{(-\alpha)^{n-1}}{(n-1)!}(a^{\dagger})^{2(n-1)} 
        \\
        &= -2\alpha a^\dagger \exp(-\alpha\left(a^{\dagger} a^{\dagger}\right)) \, .
    \end{aligned}
\end{equation}
Since $a$ annihilates the non-interacting ground state, we can replace the first term in Eq.~\eqref{eq:LL_GS_Constraint} with its commutator:
\begin{equation}
    \begin{aligned}
        \left(u a+v a^{\dagger}\right)&\exp(-\alpha\left(a^{\dagger} a^{\dagger}\right))|0\rangle=
        \\
        \left(-2u \alpha a^\dagger+v a^{\dagger}\right)&\exp(-\alpha\left(a^{\dagger} a^{\dagger}\right))|0\rangle
        = 0\,,
    \end{aligned}
\end{equation}
which immediately gives us
\begin{equation}
\label{eq:Alpha_Constraint}
\alpha = \frac{v}{2u} < \frac{1}{2} \, .
\end{equation}
The upper bound on $\alpha$ is important, which we will see when we try to normalize the state.

\section{Normalization of eigenstates}
\label{App:D}
Now we just have to normalize the interacting ground state given in Eq.~\eqref{eq:LL_GS} by finding the value of $\mathcal{N}$:
\begin{equation}
|\tilde{0}\rangle= \mathcal{N} \exp(-\alpha\left(a^{\dagger} a^{\dagger}\right))|0\rangle\,.
\end{equation}
We obtain this by the normalization constraint
\begin{equation}
\bra{\tilde{0}}\ket{\tilde{0}} = 1 = \mathcal{N}^2\bra{0} e^{-\alpha\left(aa\right)} e^{-\alpha\left(a^{\dagger} a^{\dagger}\right)}\ket{0} \, .
\end{equation}
We Taylor expand the operators to get
\begin{equation}
\sum_{n,m = 0}^{\infty} \frac{(\alpha)^{n+m}}{n!m!}(aa)^n(a^{\dagger} a^{\dagger})^m\,.
\end{equation}
Since each term is sandwiched between the ground-state, only the $n=m$ terms will survive giving us
\begin{equation}
\sum_{n,m = 0}^{\infty} \frac{(\alpha^2)^{n}}{n!n!}(aa)^n(a^{\dagger} a^{\dagger})^n \, .
\end{equation}
We make use of $(a^{\dagger} a^{\dagger})^n\ket{0} = \sqrt{2n!}\ket{n} $
to obtain
\begin{equation}
1= \mathcal{N}^2 \sum_{n= 0}^{\infty} (\alpha^2)^{n}\frac{(2n)!}{n!n!} \, .
\end{equation}
Now we have to use the formula
\begin{equation}
\frac{(2n)!}{n!n!} =\left(\begin{array}{c}
2 n \\
n
\end{array}\right)=(-4)^{n}\left(\begin{array}{c}
-\frac{1}{2} \\
n
\end{array}\right)
\end{equation}
to get
\begin{equation}
\mathcal{N}^2 \sum_{n=0}^{\infty} (-4\alpha^2)^{n}\left(\begin{array}{c}
-\frac{1}{2} \\
n
\end{array}\right) = \frac{\mathcal{N}^2}{\sqrt{1-4\alpha^2}} = 1 \,.
\end{equation}
This we get from the binomial expansion of a fractional power ($-0.5$ in this case).

We see that this series converges to a finite real value only for 
\begin{equation}
\alpha^2 < \frac{1}{4} \Rightarrow |\alpha| < \frac{1}{2} \, ,
\end{equation}
which is indeed satisfied by Eq.~\eqref{eq:Alpha_Constraint}.
Hence, we get our normalization factor as
\begin{equation}
\mathcal{N} = \sqrt{\sqrt{1-4\alpha^2}} \, .
\end{equation}

\nocite{*}

\bibliography{main.bib}

\end{document}